\documentclass[fleqn,twoside]{article}
\usepackage[headings]{espcrc2}

\usepackage{epsfig}
\usepackage[figuresright]{rotating}

\newcommand{\As}{\mbox{$\alpha_{\rm{s}}$}}
\newcommand{\Asz}{\mbox{$\alpha_{\rm{s}(M_Z)}$}}

\newcommand{\AmS}{{\protect\the\textfont2
  A\kern-.1667em\lower.5ex\hbox{M}\kern-.125emS}}

\title{Jets and Measurements of \As}

\author{P. J. Bussey\address{Department of Physics and Astronomy,\\ 
        Faculty of Physical Sciencies, \\ 
        University of Glasgow, Glasgow G12 8QQ, UK\\[2mm]
        For the H1 and ZEUS Collaborations.}
\thanks{Royal Society of Edinburgh/ Scottish Executive Support Research Fellow}
}
\begin{document}

\begin{abstract}
A survey is presented of recent measurements in jet physics, 
and improved determinations of the QCD coupling constant \As\ 
that these have made possible.
\vspace{1pc}
\end{abstract}

\maketitle

\section{INTRODUCTION}

Following the first realisation that hadronic particles consist of
charged fermions known as quarks, held together by bosonic particles
known as gluons, it was learnt that the quarks and gluons
(``partons'') can never be observed on their own.  Any attempt to
isolate a quark, for example by knocking it violently out of a hadron
in an energetic collision process, does not result in an observable
quark (or gluon).  Instead, a shower of elementary particles is seen,
referred to as a jet, containing the energy of the original quark or
gluon.  The gluons couple to the quarks by a running coupling constant
\As. 

In this account, we first make some brief notes regarding
QCD processes. There follows a survey of recent experimental
results regarding jet measurements and what has been learned from
them.  In particular, we then examine some recent measurements of 
\As\ and attempt to make some comparisons.

\section{STATUS OF CALCULATIONS}

\begin{figure}
\centerline{
\epsfig{file=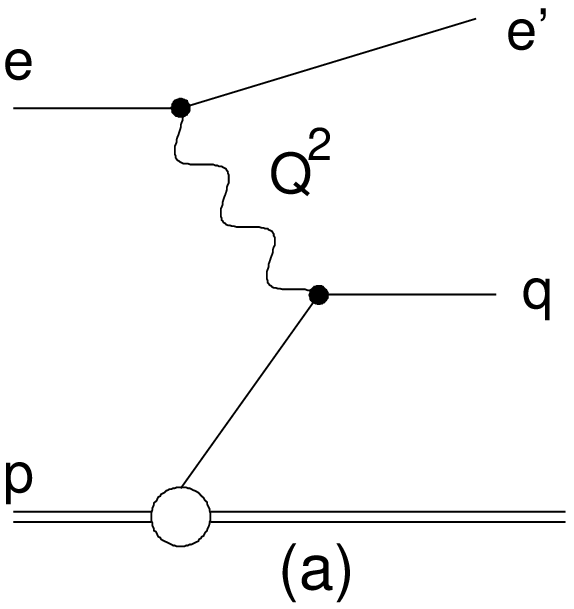,width=0.29\columnwidth}
\hspace*{0.05\columnwidth} 
\epsfig{file=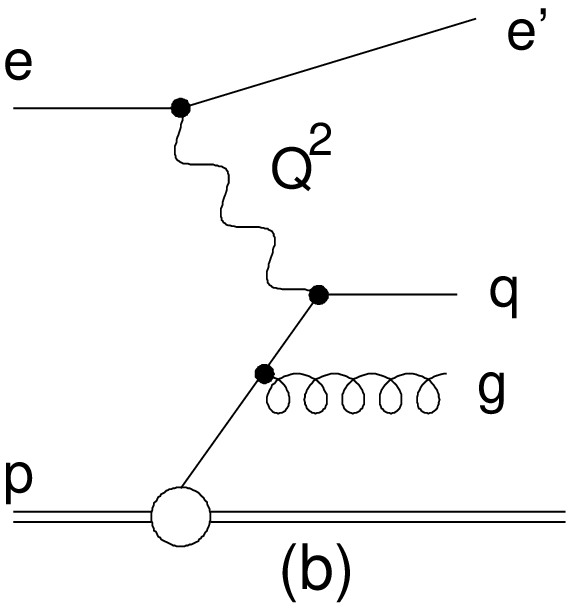,width=0.29\columnwidth} 
\hspace*{0.05\columnwidth} 
\epsfig{file=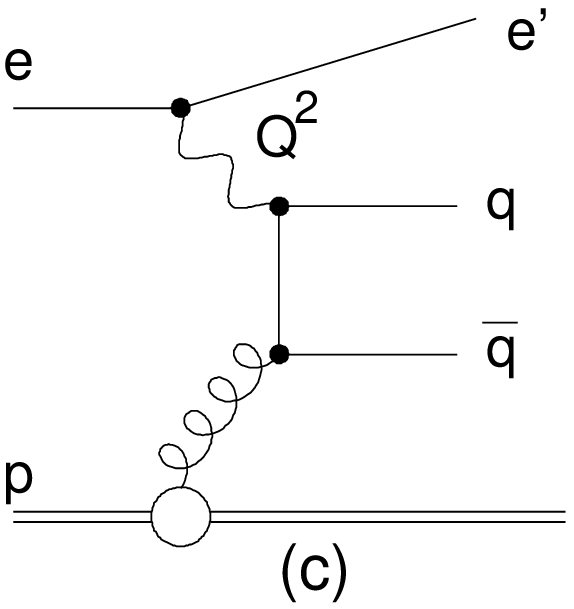,width=0.29\columnwidth} 
}
\caption{Examples of hard QCD processes at order zero and unity 
in \As.}
\label{sl6a}
\end{figure}

The most basic diagram in deep inelastic scattering (DIS) consists of
an incoming lepton exchanging a virtual boson (photon or electroweak)
with a quark in an incoming hadron (Fig.\ \ref{sl6a}(a)).  The quark
is ejected to give rise to an observable jet.  The coupling of the
virtual boson to the lepton and the parton is electromagnetic or
electroweak, and \As\ is not directly involved.

Higher order processes involve the radiation of gluons at various
points.  The outgoing quark may radiate a gluon, or an incoming
quark may do so, in which case the exchanged boson may interact
with it by means of a quark pair, a process known as boson gluon fusion.
These processes are illustrated in Figure \ref{sl6a}(b), (c)  and are of
order unity in \As.  Processes of higher order in \As\  have also been
studied and will be discussed below. 

At present \cite{GZ}, all the relevant QCD processes at lowest order
(LO) in \As\ have been calculated, and also all the next-to-leading
order (NLO) processes involving two partons entering and two leaving.
Most of the NLO processes involving the radiation of a third particle
have also been calculated, but little has been done at higher
order. The NLO calculations, as we shall see, are in good agreement with
the relevant measurements at HERA.  However theory errors are beginning to
become dominant in the \As\ determinations, raising the issue of how to obtain 
increased theoretical accuracy.

\section{EXPERIMENTAL METHODS}

The experimental methodology for measuring jet cross sections is by
now, in principle, well established.  It is taken that the jet
corresponding to a quark or gluon radiated in a hard process, namely at high
transverse momentum $p_T$, will comprise particles that are in a
definable sense close together.  Two main approaches are now adopted:
to define the closeness in terms of a cone angle in $\eta, \phi$
space, where $\eta$ is pseudorapidity and $\phi$ is angle, or to use a
clustering method.  With cones, a cone radius is specified and
particles are accepted if they are all found in one cone.  Subtleties
arise in connection with the basic cone search, and in cases where
jets overlap.  With clustering techniques a parameter is defined
involving the transverse energies of two objects and their angular
separation in $\eta, \phi$, and clustering is performed iteratively
between objects (including already clustered objects) whose clustering
parameter is less than a given value.

The complicating issues here arise in connection with the comparison
with theory. In theoretical calculations at a given order we will just
have distributions of partons, and it will be desired to apply the
same jet algorithm to these to determine the jet properties at the
parton level.  If only one parton were to be involved there is no
problem.  However this will not be the case owing to the radiation of
extra partons, and these processes tend to have infra-red and
collinear divergences; the jet algorithm must not be disturbed by
these if we are to have a definable correspondence between the parton
jets and the hadronic jets, with just a hadronisation correction to be
evaluated and applied.

The effects in question tend to be small in magnitude and were ignored
for many years in the context of cone jets, but this is no longer
acceptable if we are dealing with high order processes or precision
measurements. Clustering approaches are much less problematic,
and have become standard practice at HERA; however with many-jet events
the well-defined radius of a cone can give advantages.
The mid-point algorithm, which performs a second iteration in its cone
search, using seeds midway between the jets already
found, is helpful but not perfect.  Recent developments here include
SISCone (Seedless Infrared Safe cone) and the so-called ``anti-kT''
algorithm.  These approaches should serve well at the LHC.
For a more detailed account of these issues, the reader should consult
the account here by C. Soyez \cite{soyez}. 
 
The experimental approach, then, is to use a jet finder that will give
an adequate relationship between experiment and theory, and 
check that the basic kinematic distributions of jet quantities agree with
the QCD calculation of choice.  It will also be necessary to use 
a trusted set of parton densities (PDFs) in the proton.  Discrepancies
may indicate the need for more accurate PDFs.  If all is well, the
data may be used to extract a value of \As, provided that the process studied
involves diagrams that are of at least order unity in \As. A high-order
process would give greater sensitivity but may well not have been calculated.
These tasks, of course, may not be completely separable.

\begin{figure}[t!]
\begin{center}
\epsfig{file=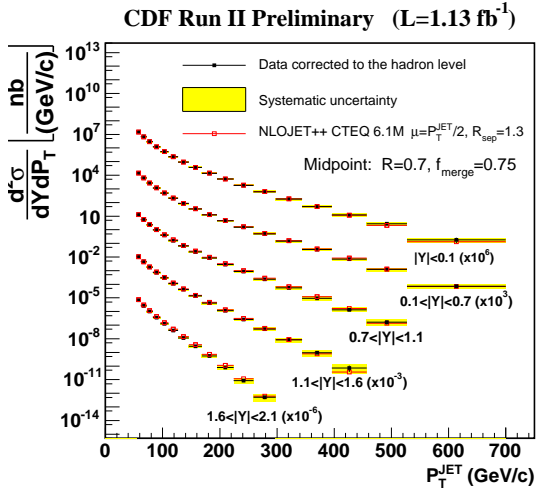,width=0.95\columnwidth,%
bbllx=20pt,bblly=0pt,bburx=521pt,bbury=475pt,clip=true}
\\[4mm]
\hspace*{0.05\columnwidth}\epsfig{file=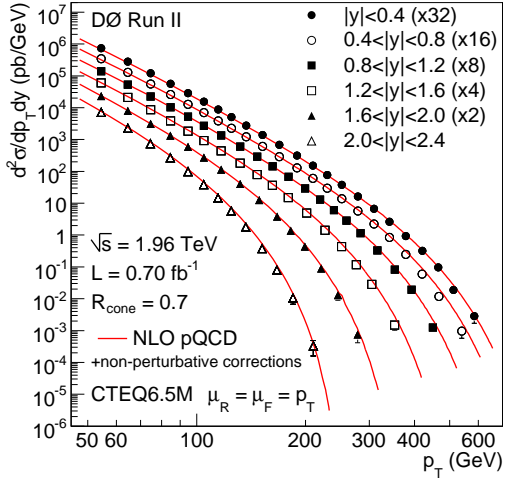,width=0.9\columnwidth}\\[-4mm] 
\end{center}
~\\[-15mm]
\caption{Jet distributions as a function of transverse energy
from CDF and D0, using the mid-point algorithm with cone radius 0.7. }
\label{CDFD0}
\end{figure}

\section{TEVATRON RESULTS}
The highest jet energies at present are achieved at the Tevatron
operating $p\bar{p}$ collisions at a centre of mass energy of 1.96
TeV.  Both CDF and D0 have made extensive studies in the field of jets
and their properties, and we show here some of their recent results
\cite{CDFjets,D0jets}. Figure \ref{CDFD0} shows inclusive jet
distributions as function of transverse energy for different rapidity
ranges. Comparison is made with an NLO QCD calculation (NLOjet) and
excellent agreement is observed, both in the shape of the
distributions and in the absolute magnitude, in all ranges.  The
agreement in magnitude amounts largely to a confirmation of the CTEQ
PDFs used, while the agreement in $E_T$ shape points to the accuracy
of the QCD calculations.

Both collaborations have also presented cross sections for prompt
photon production, which tests different perspectives on QCD.  CDF
find good agreement with theory (JETPHOX) in their results, while D0
find some discrepancies with theory in their photon plus jet cross
sections.  The cause for these is not at present clear.  CDF's dijet
mass spectrum, measured up to 1400 GeV/$c^2$, again shows excellent
agreement with NLO theory (NLOjet + CTEQ6.1M) with no signs of any
additional structural features.

\begin{figure}
\vspace*{5mm}
\begin{center}
\epsfig{file=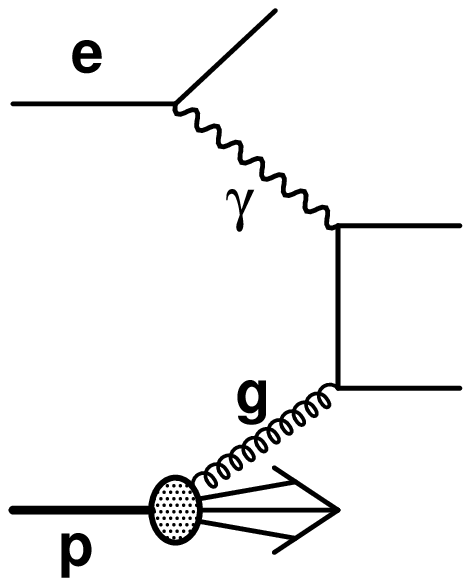,width=0.3\columnwidth}
\hspace*{0.1\columnwidth}
\epsfig{file=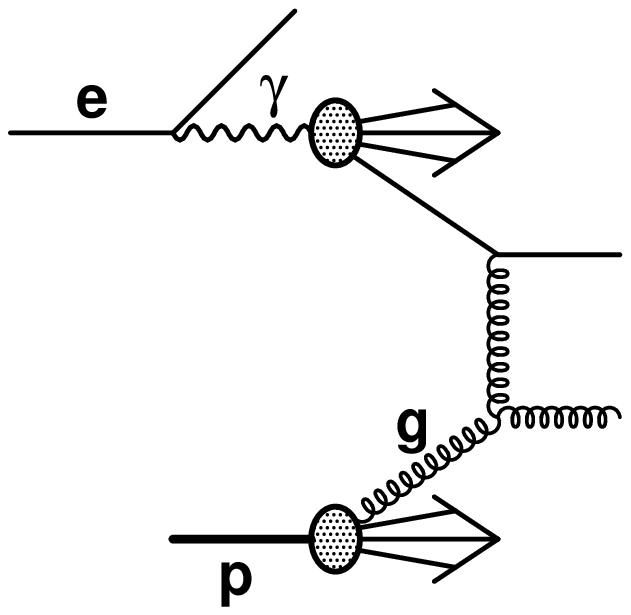,width=0.4\columnwidth}\\[-5mm]
\end{center}
~\\[-15mm]
\caption{Direct and resolved processes at LO in photoproduction.}
\label{sl11}
\end{figure}

\section{JETS IN PHOTOPRODUCTION}

Figure \ref{sl11} indicates the two main categories of process in hard
photoproduction, at lowest order in \As.  In direct processes the
photon couples directly to a high-$p_T$ quark line, while in resolved
processes, the photon interacts through a hadronic intermediate state
to which it couples non-perturbatively; this state represents the ``hadronic
structure of the photon'' and the hard scatter, describable 
perturbatively in QCD, takes place with one of the partons in this
structure.  The distinction between direct and resolved processes can
be made also at higher order, but an arbitrary division must be made
between them; only quark pairs with $p_T$ above the so-called
factorisation scale will be considered hard, while quarks at lower
$p_T$ values will be considered as constituting part of the hadronic
structure of the photon.

The fraction $x_\gamma$ of the photon energy that is taken by the jet
system, for two or more jets, can be calculated from the kinematics of
the jets.  For a direct process it should ideally take the value of
unity, but owing to hadronic effects and soft radiation its observed
value is less than this.  Direct-dominated and resolved-dominated
processes can largely be separated experimentally by taking $x_\gamma$
above and below a value of approximately 0.75.

\begin{figure}
\begin{center}
\epsfig{file=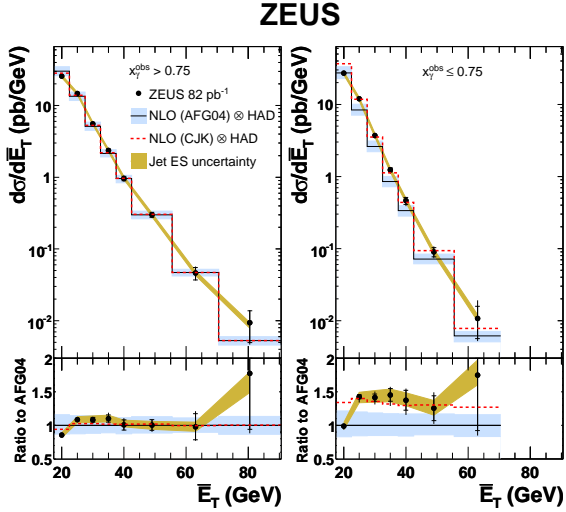,width=1.0\columnwidth}
\end{center}
\caption{Distribution of the mean $E_T$ of photoproduced dijets for 
direct-dominated and resolved-dominated event samples.
}
\label{sl12}
\end{figure}

\begin{figure}
\begin{center}
\epsfig{file=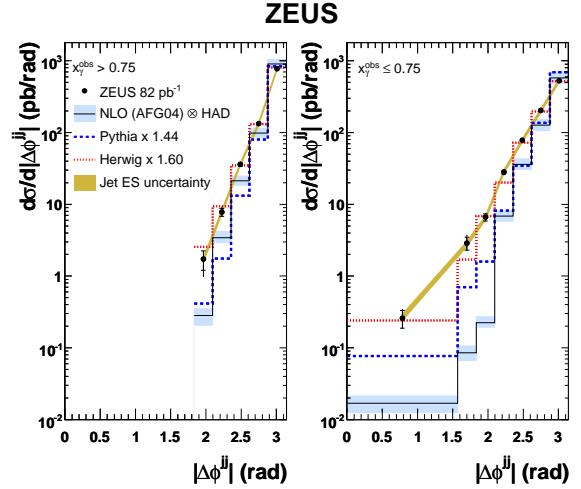,width=1.0\columnwidth}
\end{center}
\caption{
Azimuthal angle between the jets in the analysis of Figure \ref{sl12},
compared with different theoretical approaches, for direct-dominated and
resolved-dominated event samples.
}
\label{sl13}
\end{figure}

ZEUS have published improved measurements of dijet production at HERA
\cite{Zphpdij}.  The shape of the $E_T$ distribution is well described
over more than three orders of magnitude (Fig. \ref{sl12}), while the
predictions using two different hadronic models of the photon
\cite{AFG,CJK} differ slightly for the resolved events; of course
there is little difference between the photon models for the direct
events, where the excellent agreement between theory and experiment
confirms both the general theory and the choice of proton PDFs (CTEQ).
The azimuthal angle between the jets, which at LO should be
$180^\circ$, is sensitive to summed higher order effects including
soft radiation. The hadronising Monte-Carlo HERWIG is found to model this distribution better than
the standard parameters of PYTHIA, although the latter can in fact be
tuned.  (Fig. \ref{sl13})

\begin{figure}[t]
\epsfig{file=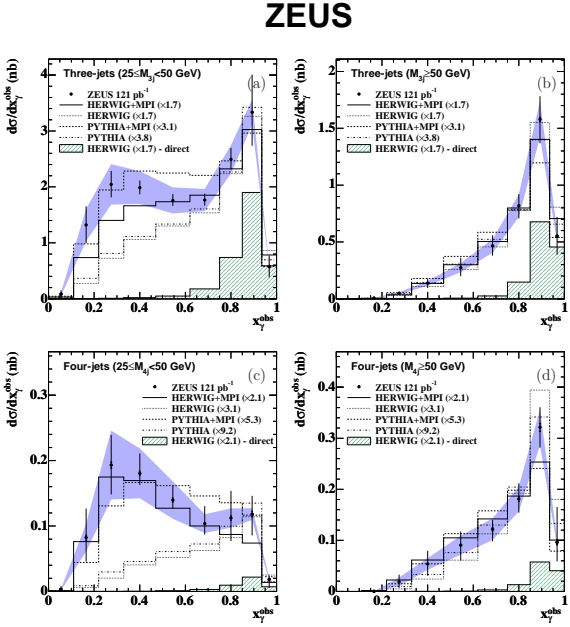,width=1.0\columnwidth}
\\[-8mm]
\caption{
Distributions in the observed value of $x_\gamma$ for three- and four-jet
events for jet masses below and above 50 GeV. 
}
\label{sl15}
\end{figure}

ZEUS have also studied multi-jet distributions with a view to checking
the accuracy of higher order calculations, and also to ascertain
whether a multi-parton description of the resolved photon-proton
scatter should be employed. The distributions in $E_T$ of the jets in
three-jet and four-jet events, in various $\eta$ intervals, are found
to be satisfactory.  There are some indications that multiple parton
interactions (MPI) improve the description.  However clearer
indications come from the $x_\gamma$ distributions (Fig. \ref{sl15}).
For high masses of the jet system, direct processes dominate, but for
low masses there is a substantial resolved component which is not well
described without the MPI option to PYTHIA or HERWIG.  Even here the
agreement is not perfect, but at least the magnitude is well given.
The effect is more pronounced in the four jet events, but is evident
in the three-jet sample.

\begin{figure}[t]
\epsfig{file=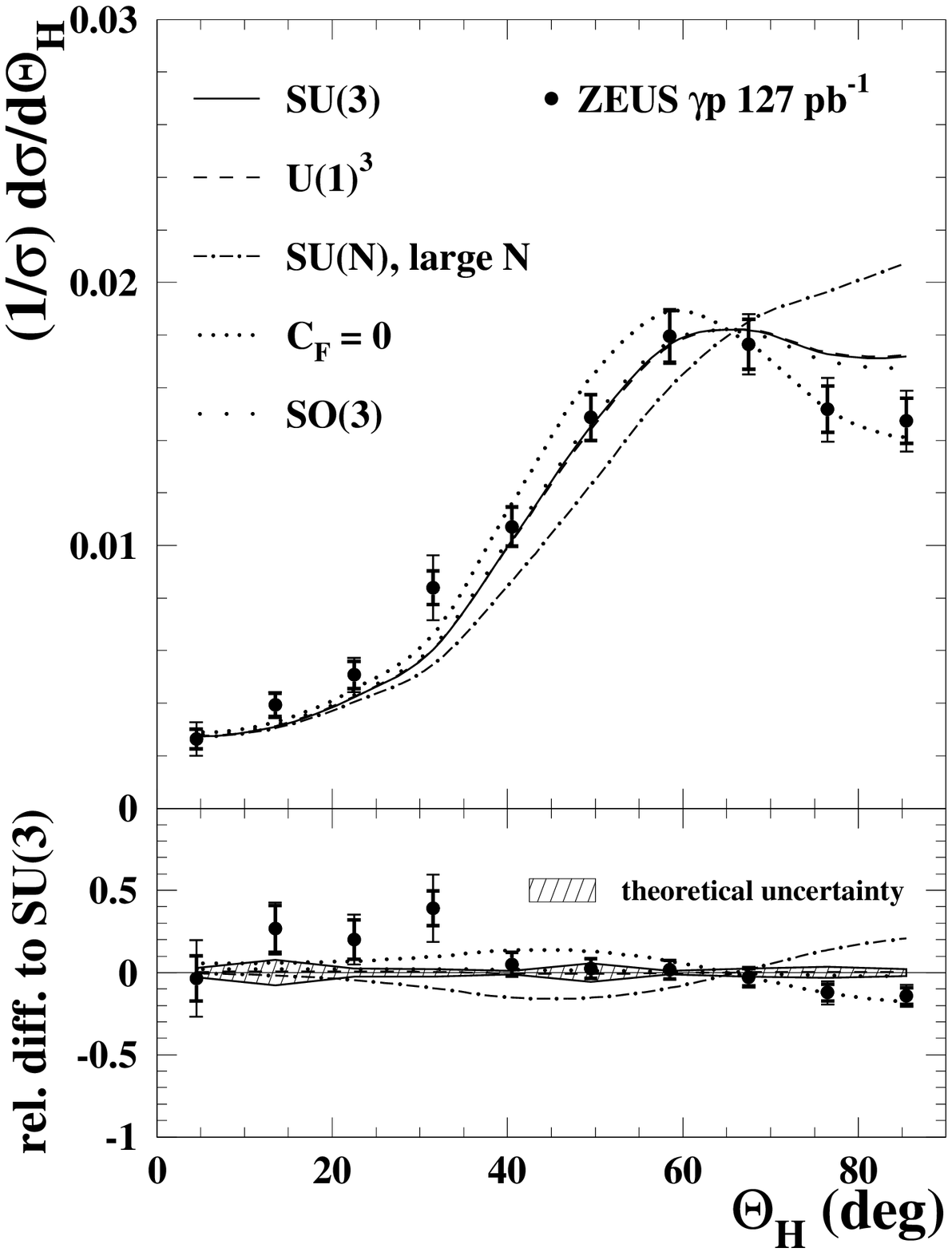,width=0.45\columnwidth,%
bbllx=30pt,bblly=0pt,bburx=455pt,bbury=540pt,clip=true}
\hspace*{0.05\columnwidth}
\epsfig{file=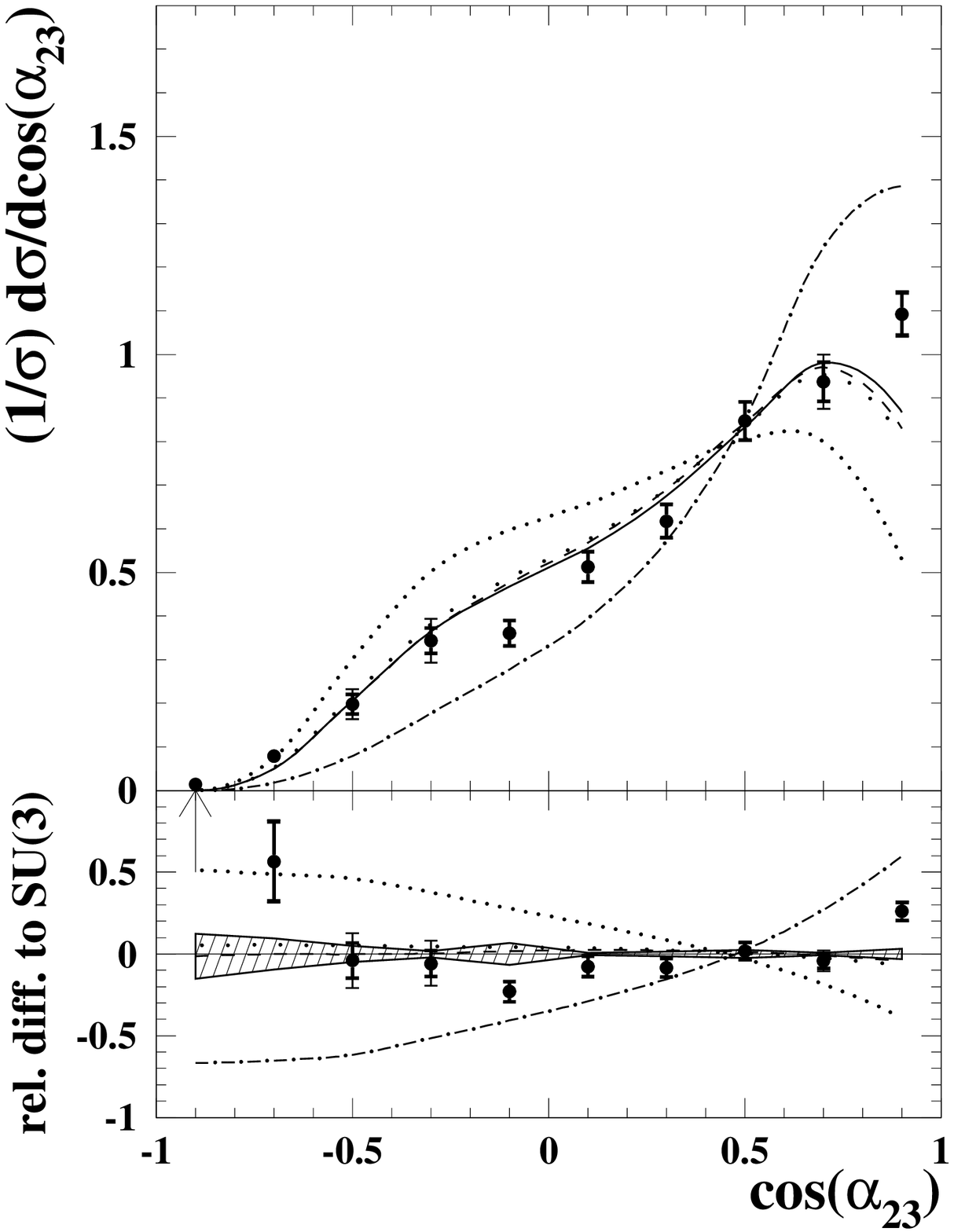,width=0.45\columnwidth,
bbllx=30pt,bblly=0pt,bburx=455pt,bbury=540pt,clip=true}
\\[1mm]
\epsfig{file=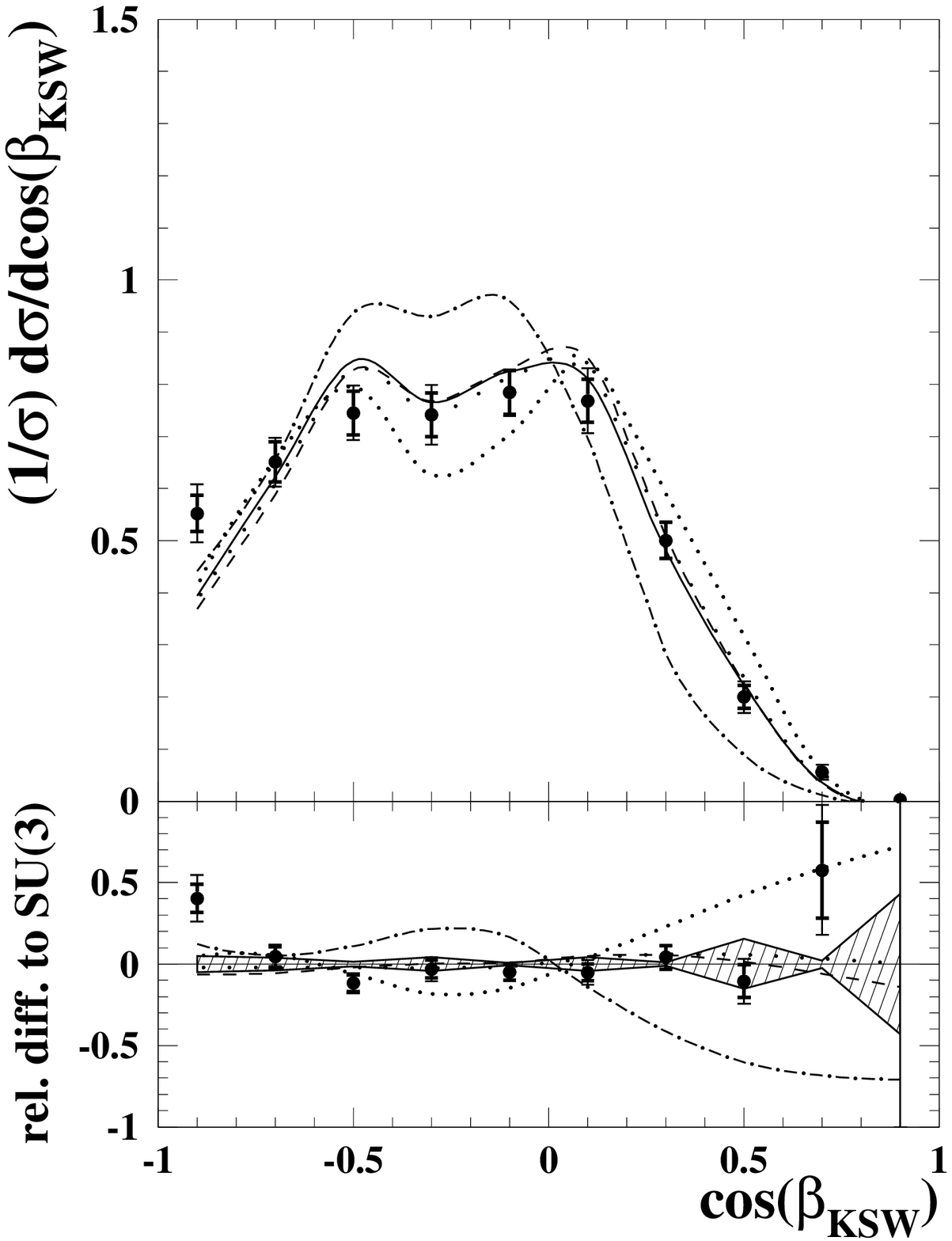,width=0.45\columnwidth,
bbllx=30pt,bblly=0pt,bburx=455pt,bbury=540pt,clip=true}\\[-5mm]
\hspace*{0.54\columnwidth}
\caption{Distribution of angular parameters of the three-jet
system with predictions from standard QCD (solid line) and several variant
theories.
}
\label{sl19}
\end{figure}

As a further study in photoproduction, ZEUS have used the angular
correlations in three-jet events to compare the predictions of
standard QCD with those from various alternative models
\cite{Z3jet}. The following parameters can be defined:
\begin{itemize}
\item $\theta_H$ = angle between plane of highest energy jet and beam, and plane of two lowest energy jets.
\item $\alpha_{23}$ = angle between two lowest energy jets.
\item $\beta_{KSW}$ = K\"orner-Schierholz-Willrodt angle (based on cross products of jet vectors).
\item $\eta^{\rm{jet}}_{\rm{\max}}$ = maximum pseudorapidity of the three jets.
\end{itemize}
Distributions of these variables can then be compared with the
predictions from different variant models.  QCD makes use of the
symmetry group SU(3); we can compare this with SU(N) with large N,
U(1)$^3$ (no triple gluon coupling), and SO(3).  The results,
unsurprisingly but gratifyingly, favour standard QCD. Figure
\ref{sl19} illustrates some of the distributions. In each case,
standard QCD fits the data well, while the other possibilities all
fail in some regions of some of the plots.

\section{EFFECT OF POLARISED BEAMS}

Turning to deep inelastic scattering, we note that HERA II was able to
deliver longitudinally polarised electron and positron beams.  In the
electroweak sector of the standard model, the charged-current (CC) jet
cross sections are sensitive both to the polarisation and to the
nature of the lepton beam.  Since CC DIS converts the incoming lepton
into a neutrino, the signature for this process is at least one
$p_T$-imbalanced hard jet in the detector.  The cross sections become
measurable only at high values of the virtual boson mass squared $Q^2$.  As
in all the HERA measurements, a $k_T$ cluster algorithm was used for
jet identification.

\begin{figure}[t]
\epsfig{file=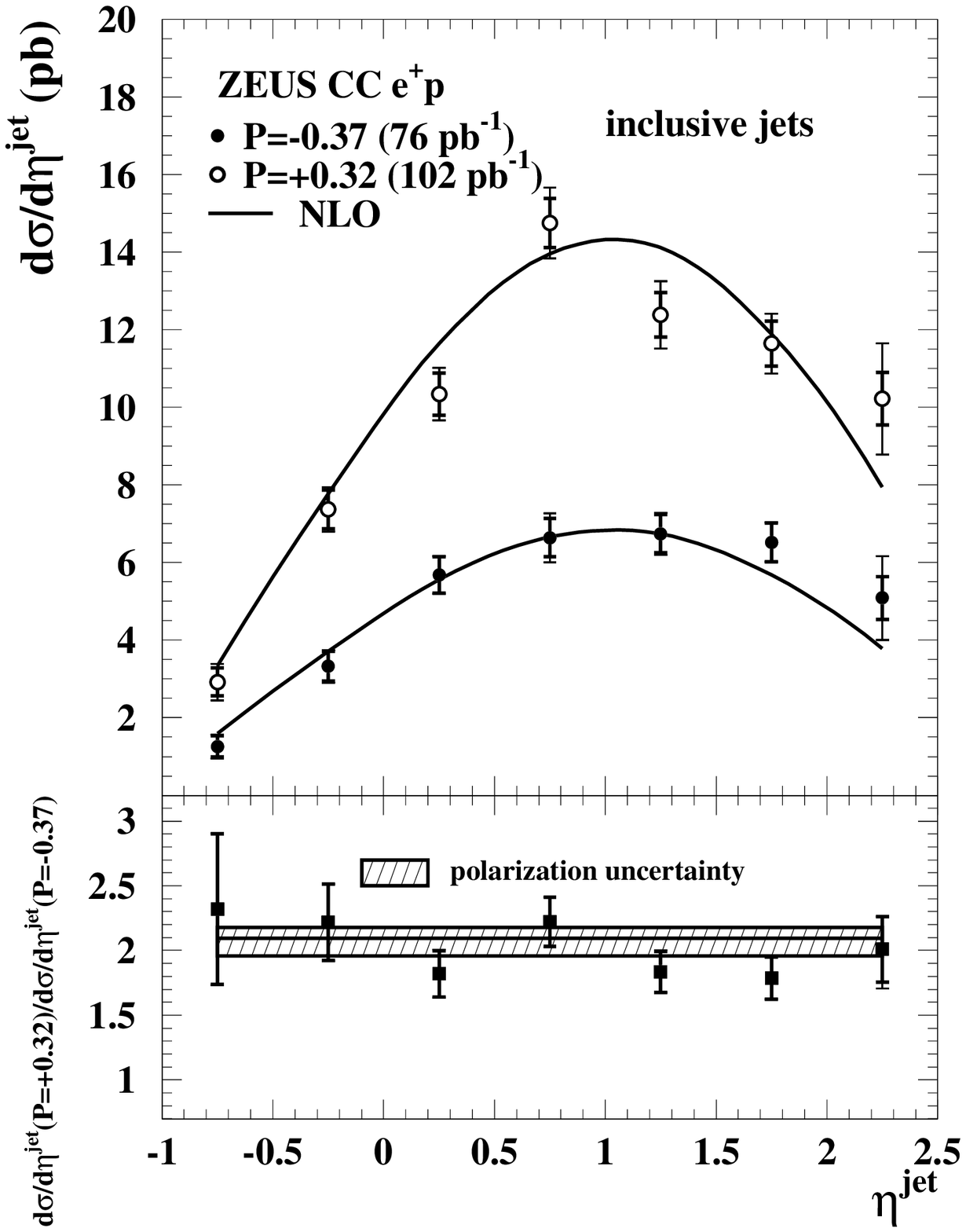,width=0.45\columnwidth,%
bbllx=35pt,bblly=0pt,bburx=465pt,bbury=546pt,clip=true}
\hspace*{0.05\columnwidth}
\epsfig{file=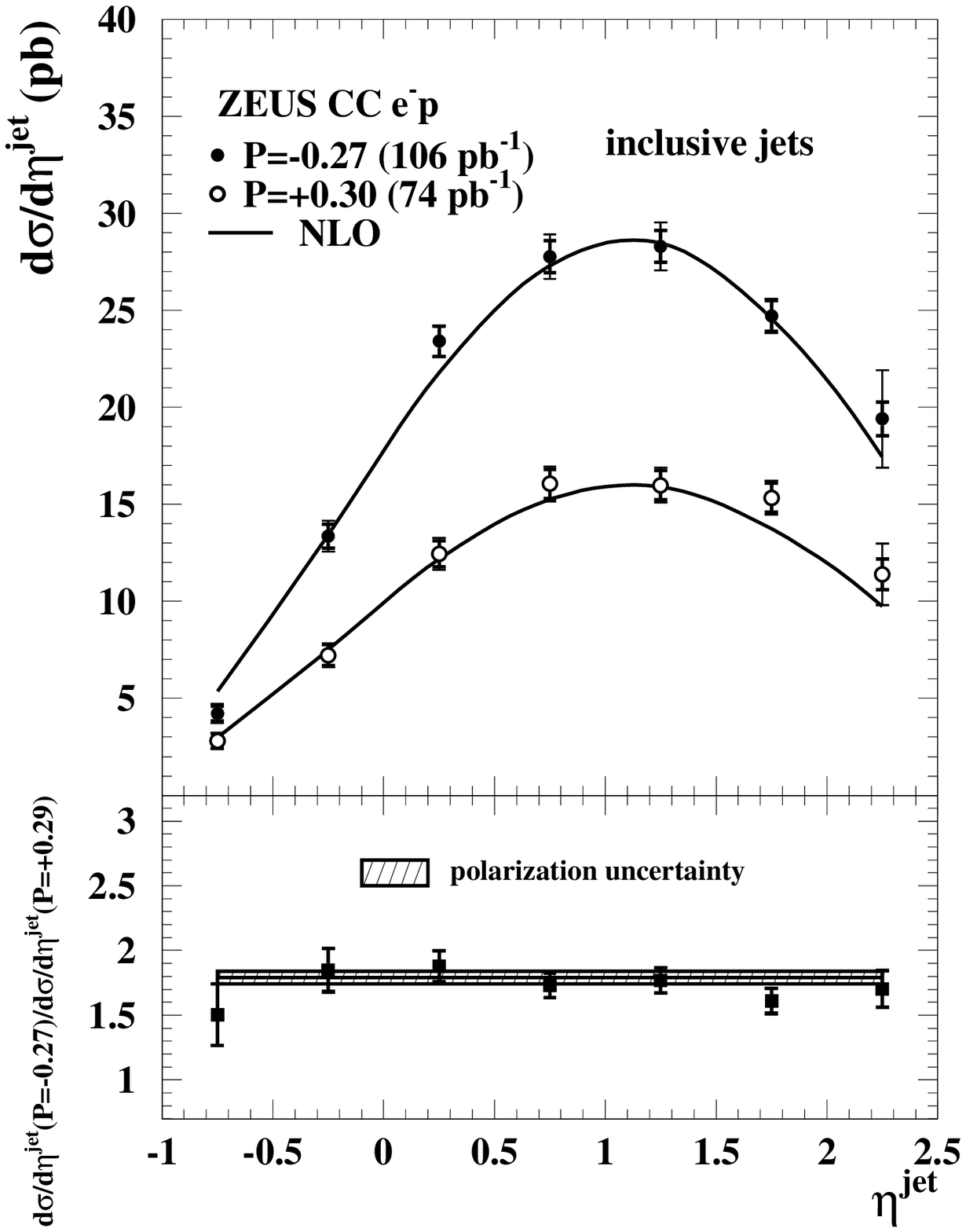,width=0.45\columnwidth,%
bbllx=35pt,bblly=0pt,bburx=465pt,bbury=546pt,clip=true}
\caption{Cross section measured by ZEUS for inclusive jets in charged-current
DIS for positron and electron beams, and opposite longitudinal
electron/positron polarisations, compared to SM theory at NLO (MEPJET)
as a function of jet pseudorapidity. }
\label{sl22}
\end{figure} 

Figure \ref{sl22} illustrates the results obtained \cite{ZCC}.  It is
clear that the cross sections are much higher for positive
polarisations than negative for a positron beam, and vice versa for
an electron beam.  Agreement with theory, using a ZEUS
parameterisation of the proton PDFs, is very good, and was confirmed using
the CTEQ6 and MRST2001 PDF sets.  Comparable results are obtained for the
cross sections as a function of $E_T$ of the jets, and for the cross
sections evaluated for zero beam polarisation.  Here we are
specifically testing electroweak theory in a new regime from the
measurements at LEP. 

\section{BREIT FRAME MEASUREMENTS.}

A good method for  measuring \As\ in DIS is to determine the 
cross sections for different kinematic  ranges as a function of 
$Q^2$ and make use of the DGLAP equations, which involve \As, 
to describe the variation of the cross sections with $Q^2$ and
the kinematic variables.  A fit to the entire data set enables
\As\ to be extracted.

Here, however, we concentrate on \As\ determinations which depend more
explicitly on measuring jet cross sections.  H1 have performed this
task as follows.  First, they take each event as viewed in the Breit
frame and find jets using the $k_T$ cluster method in this frame.  In
the Breit frame, the event axis is taken along the direction of the
virtual photon, such that in a simple deep inelastic scatter
(Fig. 1(a)) the quark is hit head-on and merely reverses its
direction.  Such scatters, although they are hard, have no transverse energy in
the Breit frame.  It follows that high-$E_T$ jets in this frame result
exclusively from higher order processes, namely those that involve
non-zero powers of \As.  The sensitivity to \As\ is therefore enhanced
by measuring jets in this way.

\begin{figure}[t]
\epsfig{file=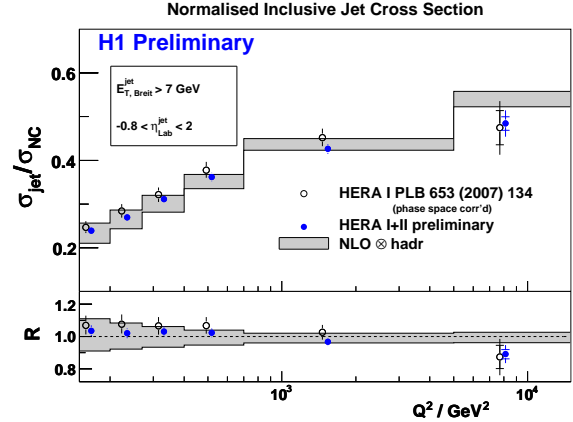,width=0.74\columnwidth,%
bbllx=100pt,bblly=70pt,bburx=541pt,bbury=670pt,clip=true,angle=-90}
\caption{Cross section for inclusive jets in the Breit frame,
with $E_T > 7$ GeV.
The values are normalised to the total neutral current cross section
at a given value of $Q^2$.
}
\label{sl25}
\end{figure} 

A laboratory rapidity cut is imposed in order to have well-defined
acceptance for the jets, and a number of different differential cross
sections are evaluated.  Figure \ref{sl25} shows normalised inclusive
jet cross sections measured under these conditions, compared to a NLO
calculation; similar distributions were obtained for two- and
three-jet final states. The normalisation approach enables some of the
experimental and theoretical uncertainties in \As\ to be reduced.  This
analysis was performed at high values of $Q^2$, which reduces some of
the theoretical errors at the cost of poorer statistics.  All the
control plots gave good fits using the NLO theory. From a
general fit to the data,
\As\ was evaluated as a function of $Q^2$.

\begin{figure}[b!]
\begin{center}
\epsfig{file=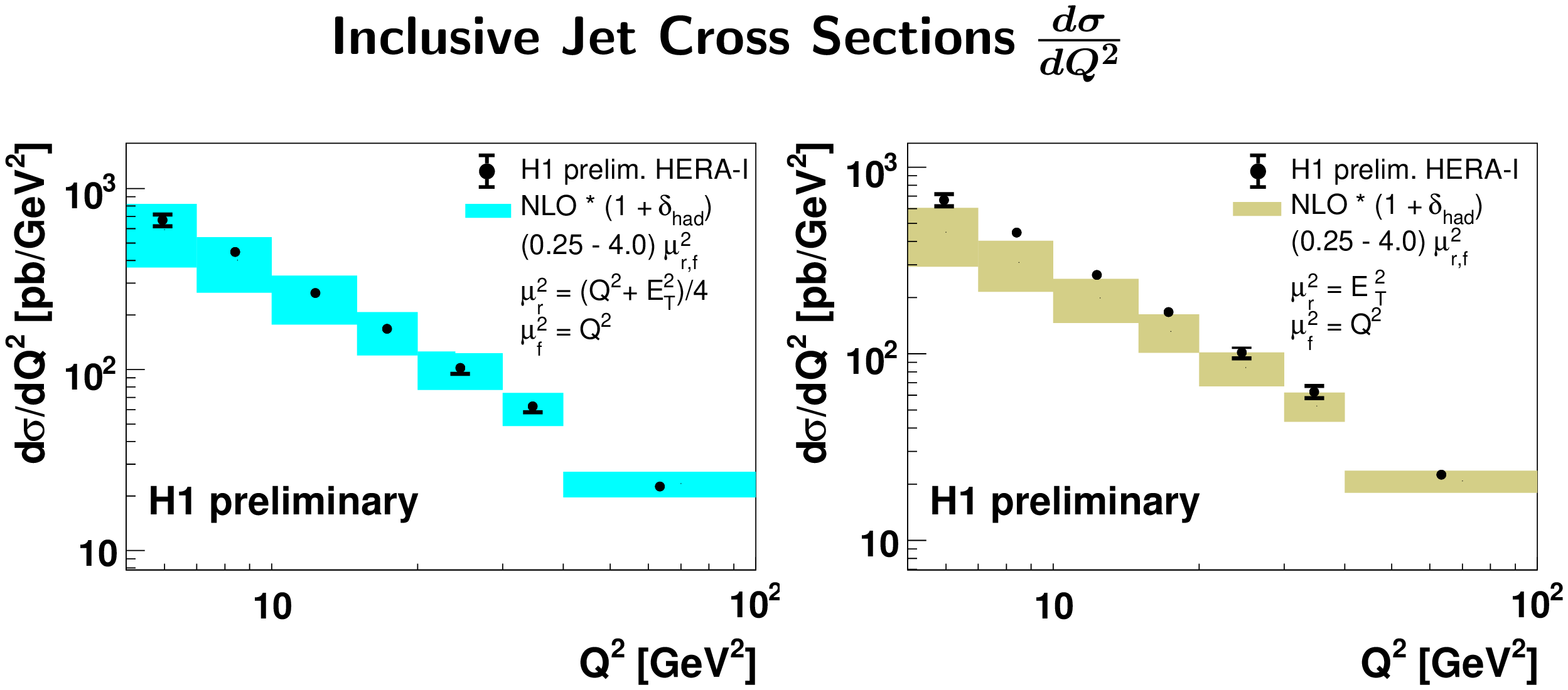,width=0.45\columnwidth,%
bbllx=60pt,bblly=80pt,bburx=390pt,bbury=850pt,clip=true,angle=-90}
\\[3mm]
\epsfig{file=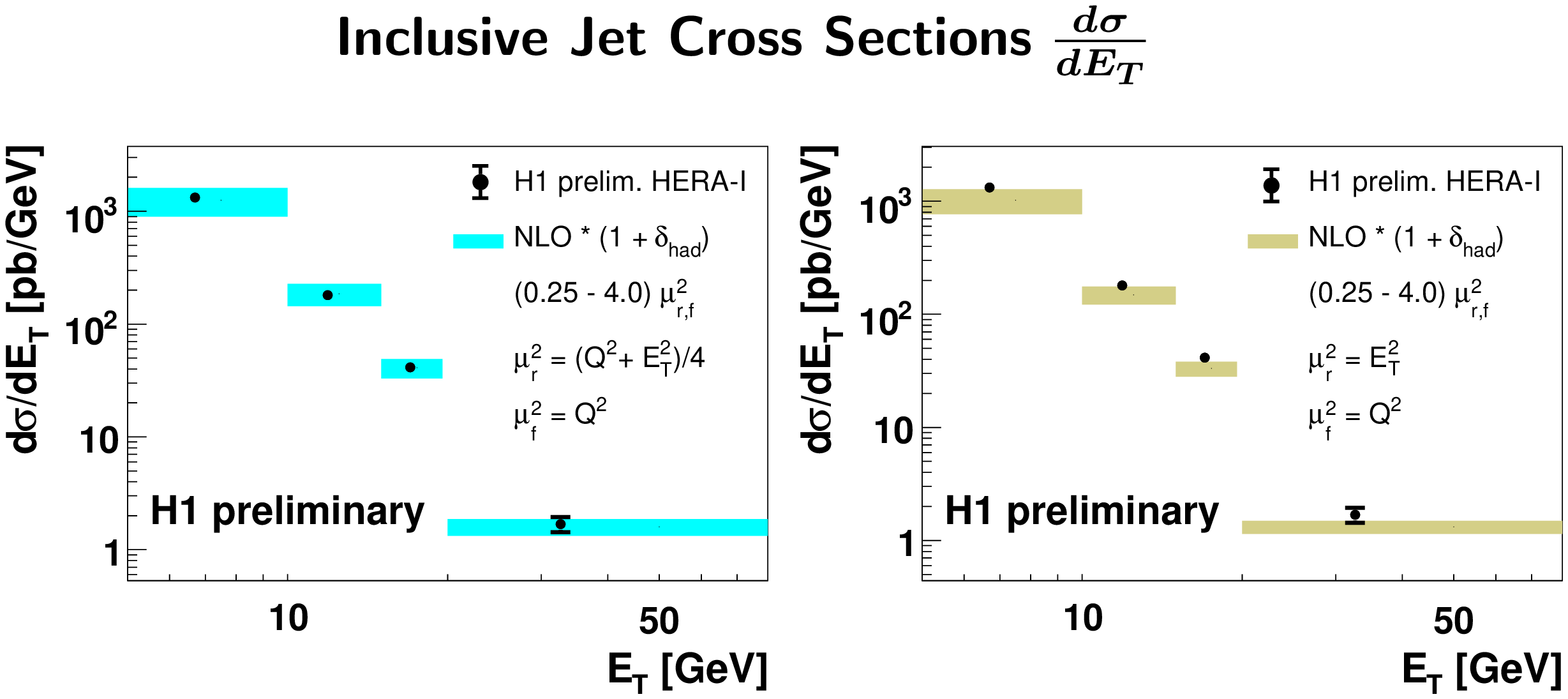,width=0.45\columnwidth,%
bbllx=60pt,bblly=80pt,bburx=390pt,bbury=850pt,clip=true,angle=-90}
\end{center}
\caption{ H1 differential cross section values 
for inclusive jets in the Breit frame
in DIS for a lower range of $Q^2$ values. Left: standard renormalisation
scale; right: variant renormalisation scale.
}
\label{sl256}
\end{figure} 

In a second analysis, H1 used a lower range of $Q^2$, giving better
statistics but, as we shall see, notably worse theoretical
uncertainties.  Events are selected in the range $5 < Q^2 < 100$
GeV$^2$ with the fractional virtual photon energy in the range $0.2 <
y < 0.7$.  Jets are selected, as before, in the Breit frame and must
have $E_T > 5$ GeV.  Detector effects are simulated and acceptances
evaluated, as before, using standard Monte Carlos. General comparison
is made with the NLOJET++ Monte Carlo predictions.

As is evident in Figure \ref{sl256}, there is a noticeable difference
between the predictions for different choices of renormalisation
scale. Both $Q^2$ and $(E_T^2 +Q^2)/4$ are plausible, but the former
value provides a good fit to the data while the latter is barely
acceptable.  The difference between the final results obtained with
these two choices must therefore be accounted as a theoretical
uncertainty.  Varying the factorisation scale likewise gives rise to a
theoretical uncertainty on the results.  The experimental systematics
must likewise be included in the normal way, and variations in the
chosen PDFs must also be taken.

For a series of $Q^2$ bands, double differential cross sections
$d^2\sigma/dQ^2dE_T$ were evaluated for a set of jet $E_T$ values.
Fits varying \As\ were carried out on these.
The results are given in the next section.

\section{RESULTS FOR \As}

Experimental values for \As\ can be determined in various ways, and they are
seen to fall with increasing values of the typical momentum scale,
such as $Q$ or an equivalent variable, demonstrating the
running nature of this coupling constant. It is conventionally quoted
at $Q^2 = M_Z^2$. At LEP, some accurate determinations have been made
using the properties of event shapes.  Dissertori et al \cite {Diss}
have used an NNLO QCD calcluation to fit ALEPH data and obtain $$ \Asz
= 0.1240 \pm 0.0008\pm 0.0010\pm 0.0011\pm 0.0029 $$ where the errors
are, in order, statistical, systematic, hadronisation and
theoretical. From an analysis of ALEPH and OPAL thrust data, and
incorporating their own theoretical model, Becher and Schwartz
\cite{BS} have obtained $$ \Asz = 0.1172 \pm 0.0010 \pm 0.0008 \pm
0.0012 \pm 0.0012 $$ Using an NNLO + NLLA ansatz, Bethke et
al. \cite{Bethke} obtain from the JADE data $$ \Asz = 0.1172 \pm
0.0006 \pm 0.0020 \pm 0.0035 \pm 0.0030. $$ These values have high
numerical precision but have potential dependencies on the theoretical
approach that is adopted.

\begin{figure}[t]
\begin{center}
\epsfig{file=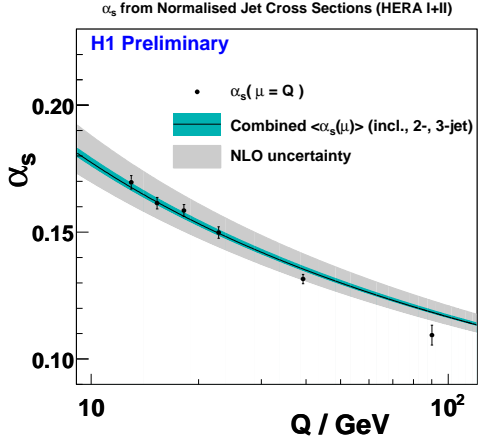,width=0.8\columnwidth,%
bbllx=0pt,bblly=20pt,bburx=570pt,bbury=650pt,clip=true,angle=-90}
\end{center}
\caption{Fitted values of \As\ from H1 data at high $Q^2$, as a function of
$Q$.  The narrow central band indicates experimental uncertainties and the
broader outer band indicates theoretical uncertainties.
}
\label{sl31}
\end{figure} 

\begin{figure}[t!]
\begin{center}
\epsfig{file=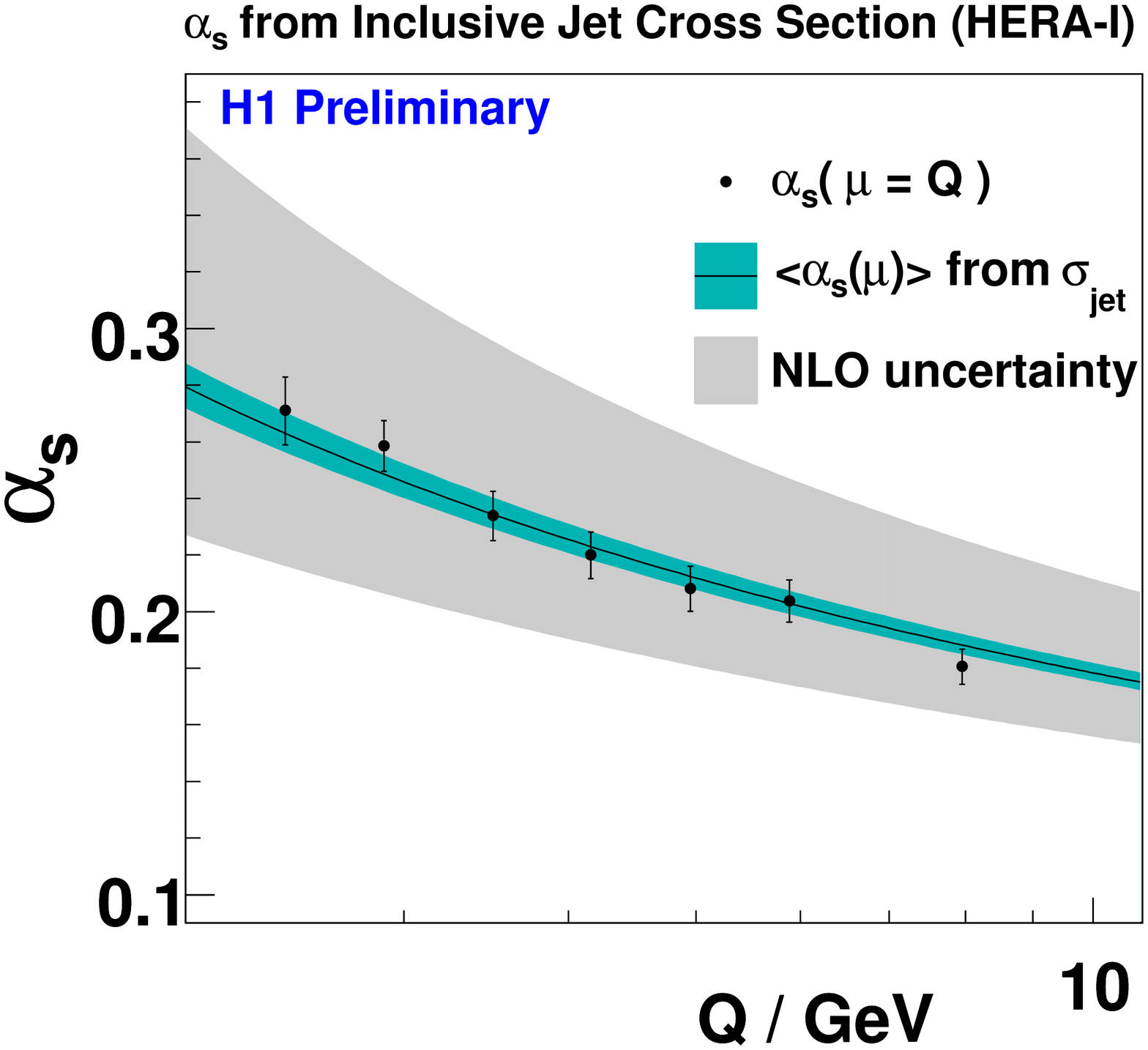,width=0.8\columnwidth,%
bbllx=30pt,bblly=30pt,bburx=570pt,bbury=650pt,clip=true,angle=-90}
\\[5mm]
\epsfig{file=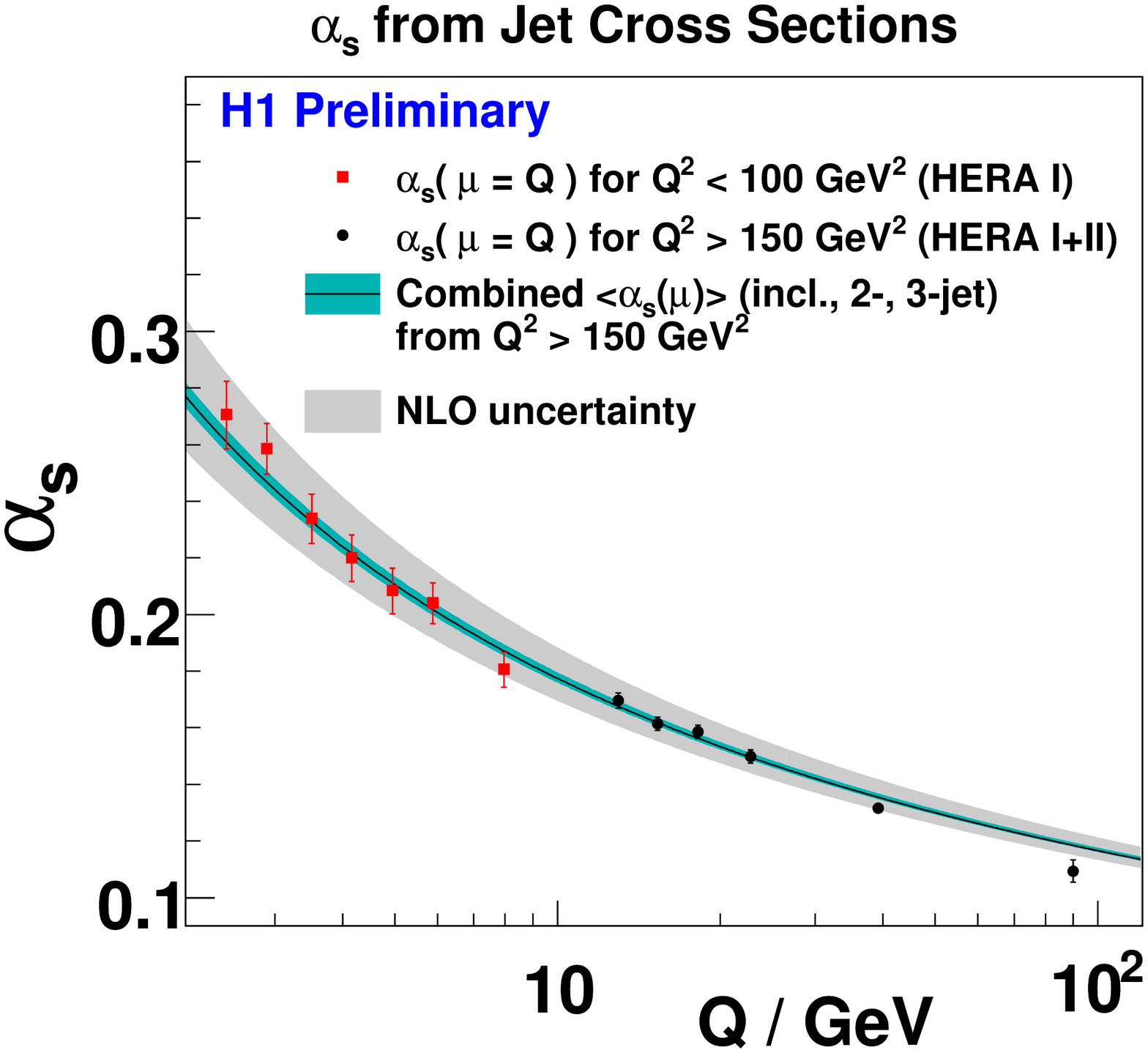,width=0.8\columnwidth,%
bbllx=30pt,bblly=30pt,bburx=570pt,bbury=650pt,clip=true,angle=-90}
\end{center}
\caption{Fitted values of \As\ from H1 data at high $Q^2$, as a function of
$Q$.  The narrow central band indicates experimental uncertainties
and the broader outer band indicates theoretical uncertainties. In
the lower plot the high-$Q$ band is extrapolated into the low-$Q$ region. }
\label{sl32}
\end{figure}

From HERA we first examine the recent H1 high-$Q^2$ results, obtained
from the data described above (Fig.\ \ref{sl31}), and representing a
clear improvement on their publication of 2007 \cite{H1incl}.
Sensitivity to \As\ is increased by measuring events with differing
numbers of jets.  Having demonstrated that \As\ runs as expected with
$Q^2$, the group perform a combined fit and obtain the value $$ \Asz
=0.1182\pm 0.0008\;^{+0.0041}_{-0.0031}\pm 0.0018. $$ The low-$Q^2$
results give a fitted value of $$ \Asz = 0.1186 \pm0.0014
\;^{+0.0132}_{-0.0101}
\pm0.0021 $$ where the three uncertainties correspond to
experiment(total), theory, and PDFs.  These results (Fig.\
\ref{sl32}, top) have a very much larger theoretical uncertainty than
the high-$Q^2$ set.  However when the values obtained are plotted with
the fitted values and certainties obtained from the high-$Q^2$ set
extended down to lower $Q^2$, the agreement is perfect, indicating the
consistency of the central values of the theoretical parameters
applied over the whole range of $Q^2$. Given the smallness of the
experimental compared to the theoretical uncertainties, the need for
NNLO calculations is becoming increasingly evident.

\begin{figure}[t!]
\epsfig{file=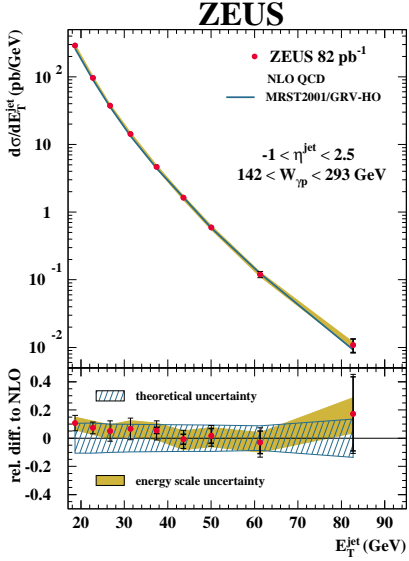,width=1.0\columnwidth}
\caption{Inclusive jet distribution measured by ZEUS in photoproduction
with $E_T > 17$ GeV, compared to NLO theory.}
\label{sl33}
\end{figure}

ZEUS have evaluated \As\ from some recent measurement of inclusive jets
in photoproduction (Fig.\ \ref{sl33}), taking the MRST2001 and GRV-HO proton
and photon structures as central values.  They obtain 
$$
\Asz = 0.1223 \pm0.0001 \pm0.0022 \pm 0.0030,$$
where the uncertainties are respectively statistical, systematic and theoretical.
This is competitive with other values and in good agreement with
them.  In DIS, H1 and ZEUS have put together their results for HERA I
inclusive jet data to produce a combined fit for \As.  Apart from the
improved statistical accuracy, this approach allows for a correct
account to be taken of systematic and theoretical effects which are
common or correlated between the two experiments.  From H1, 24 data
points at different $E_T$ are taken with $150 < Q^2 < 15000$ GeV$^2$,
while ZEUS contributes six inclusive cross sections in the range $125
< Q^2 < 100000$ GeV$^2$.  The QCD cross sections are calculated with
NLOJET++.  The factorisation scale is taken as $Q$, the
renormalisation scale is taken as $E_T$ of the jets, and the MRST2001
PDF sets are employed.  A $\chi^2$ matrix is defined and minimised
using a Hessian method.  The result is: $$\Asz = 0.1198 \pm 0.0019
(exp) \pm0.0026 (theory),$$ competitive with the LEP results.

\begin{figure}[t!]
~\\[-13mm]
\epsfig{file=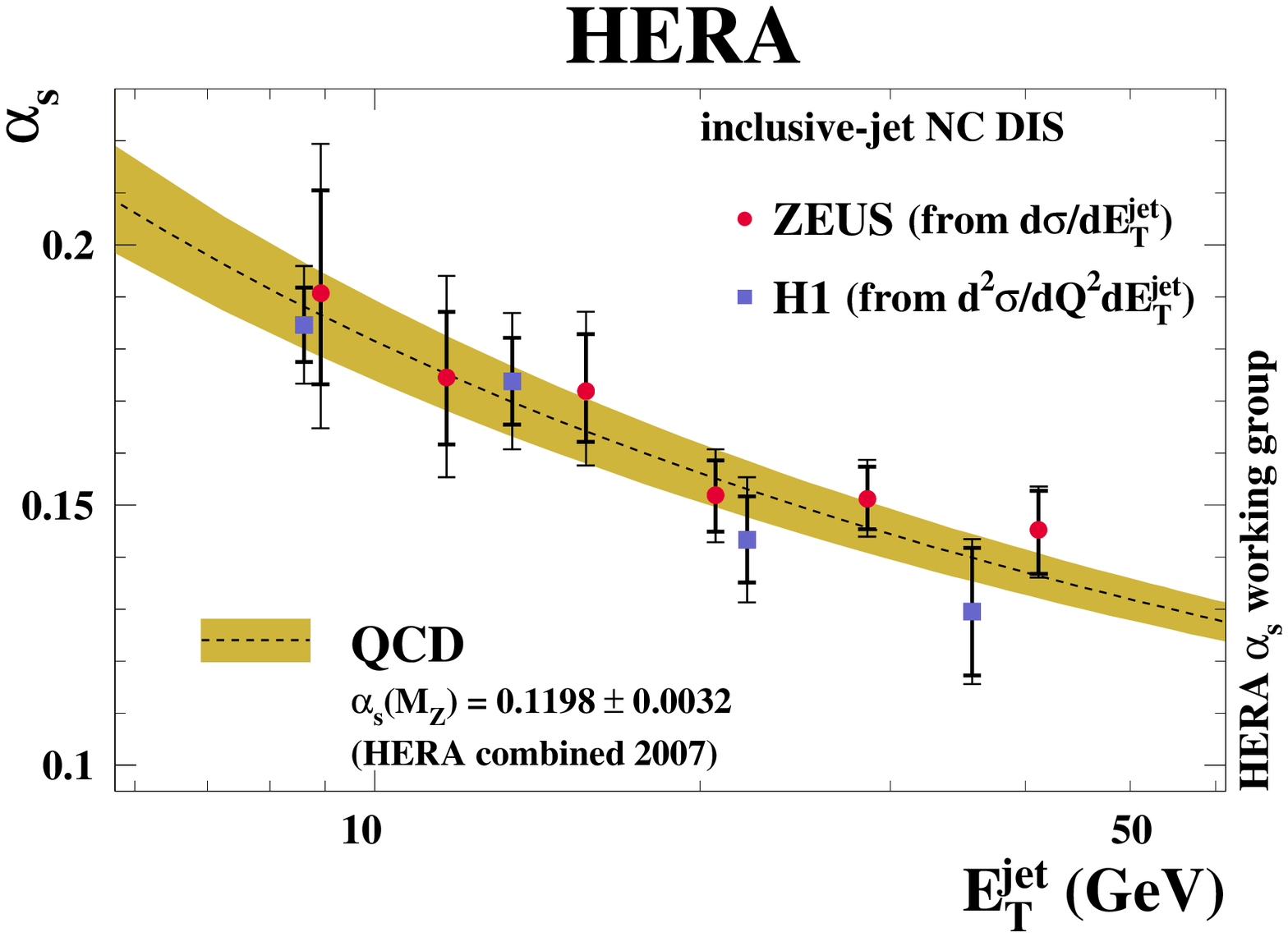,width=1.0\columnwidth}
\\[-9mm]
\epsfig{file=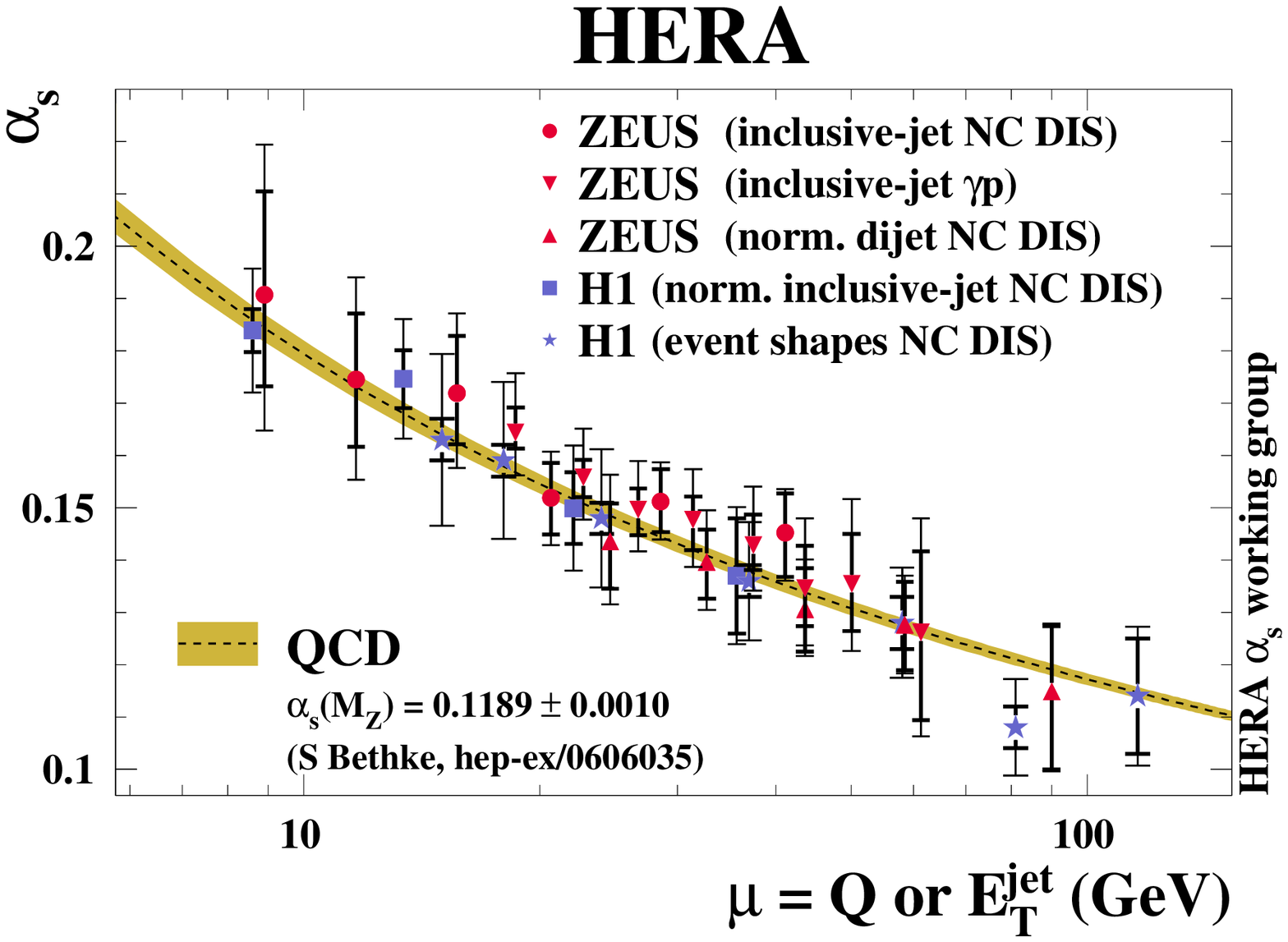,width=1.0\columnwidth}
\caption{Fitted values of \As\ from H1 and ZEUS data at high $Q^2$, 
as a function of $Q$. The top plot illustrates results from the combined
fit; the lower plot includes a wider collection.  }
\label{sl35}
\end{figure}

\begin{figure}
\begin{center}
\epsfig{file=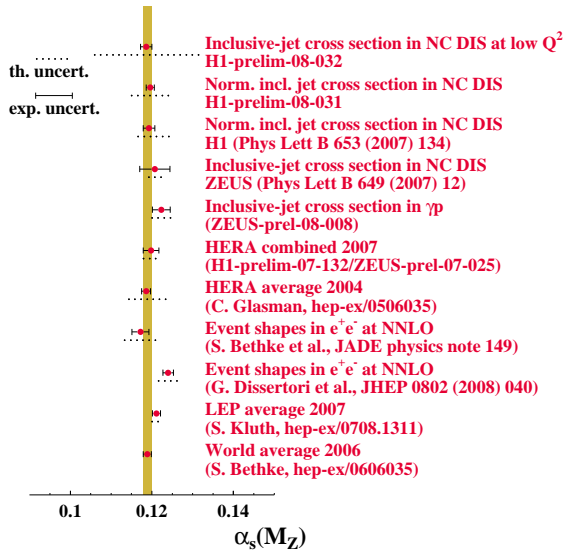,width=1.0\columnwidth,%
bbllx=0pt,bblly=0pt,bburx=525pt,bbury=525pt,clip=true}
\end{center}
\caption{Comparison of various determinations of \As\ with a world
average value.}
\label{sl36}
\end{figure}

Figure \ref{sl35} illustrates results from the combined fit, showing
the running of \As\ with jet $E_T$, and also a collected set of \As
values from a number of HERA analyses, again showing well the running
of \As\ with the relevant momentum scale.  A collection of
determinations of \As\ is shown in Figure \ref{sl36} in comparison with
a recent world average value.  It is clear that the different
approaches are in general in good agreement.  The tendency is also
evident that in order to obtain small theoretical uncertainties, which
in practice means using high momentum scales, a penalty of poorer
statistics must be paid.  This effect will become less evident
if sufficient  statistics become available to reduce the dominance of
the statistical uncertainty, as is already the case in some analyses.

\section{SUMMARY}
In summary, the study of jet distributions at HERA, as at the
Tevatron, has been a highly fruitful area to investigate a number of
aspects of Quantum Chromo-Dynamics.  The theory continues to pass all
tests, and evaluations of the coupling constant \As\ are being
performed from a number of different perspectives.  Accuracy is improving with
the accumulation of larger data samples and further developments in
the theoretical calculations that are available for the
experimentalists to use. However there is an increasing need for more NNLO
calculations. Nevertheless, this is impressive progress and should
come to a fruition when the full HERA II analyses have been completed.

~\\
\noindent
I should like to thank the organisers of the Ringberg 2008 workshop for 
all the efforts that they put in to ensure an outstandingly enjoyable and stimulating meeting.

\end{document}